\documentclass{revtex4}
\usepackage {amssymb}
\usepackage {amsmath}
\usepackage {epsfig}
\usepackage{graphicx}
\usepackage {hyperref}

\def\be {\begin{equation}}
\def\ee {\end{equation}}
\def\beq{\begin{eqnarray}}
\def\eeq{\end{eqnarray}}
\def\bea{\begin{eqnarray}}
\def\eea{\end{eqnarray}}
\def\nonu{\nonumber}
\begin{document}

\markboth{M. Horta\c{c}su and B.C.
L\"{u}tf\"{u}o\={g}lu}{Renormalization Group Analysis of a G\"
ursey Model Inspired Field Theory}

\title{Renormalization Group Analysis of a G\" ursey Model Inspired Field Theory}%

\author{M. Horta\c{c}su}
\email{hortacsu@itu.edu.tr}%

\author{B.C. L\"{u}tf\"{u}o\={g}lu}
\email{bcan@itu.edu.tr}%

\affiliation{Department of Physics, Istanbul Technical University,
Istanbul, Turkey.}%


\begin{abstract}
We  show that when a model, which is equivalent to the G\" ursey
model classically, is gauged with a SU(N) field, we get
indications of a nontrivial field theory.
\end{abstract}%

\maketitle%

\section{Introduction}

To write a field theoretical model  which has nonzero values for
the coupling constants at zeroes of the beta function of the
renormalization group is an endeavor which is still continuing in
particle physics. The  $\phi^4$ theory is a "laboratory" where
different methods in quantum field theory are first applied. After
it was shown that this model became a trivial theory when the
cutoff was removed, \cite{ba_ki_79,ba_ki_81},  it was clear that
analyzing the terms in the perturbation series was not sufficient
to decide whether one had a truly interacting theory. Work in this
field was also given by Wilson and others, \cite{wi_73,kl_06}.
Renormalization group methods, first introduced by Wilson for this
purpose \cite{wi_ko_74}, are the most commonly used technique in
studying whether one has a trivial theory or not.

\noindent Since  a nontrivial fixed point is not yet found for
QCD, there are attempts to study alternative models for this
purpose is given in \cite{gi_ja_we_04}. A very popular model is
the Nambu-Jona-Lasinio model, hereafter NJL, \cite{na_jo_61}. This
model is written in terms of spinor fields only, and is used as an
effective theory extensively in high energy physics
\cite{mu_87,mi_93}. The NJL model was also shown to be trivial
\cite{ko_ko_94,zi_89}. Recent attempts to gauge this model to
obtain a nontrivial theory are given in references
\cite{ha_ki_ku_na_94,ao_mo_su_te_to_99,ao_mo_su_te_to_00,ku_te_99,ko_ta_ya_93}.
Both functional and diagram summing methods were used in these
papers.  Exact renormalization group methods proposed by Wilson
and Polchinski, \cite{wi_ko_74,po_84}, are often employed for this
purpose.  A very recent paper on this method is given by Sonoda
\cite{so_07} .

\noindent Another model, which uses only spinors is the G\" ursey
model \cite{gu_56}. We have worked on different forms of the G\"
ursey model \cite{ho_lu_06,ho_ta_07,ho_lu_ta_hepth_06}. Our
starting point was both our earlier work,
\cite{ak_ar_du_ho_ka_pa_82-34,ak_ar_du_ho_ka_pa_82-41,ak_ar_ho_pa_83,ar_ho_83,ar_ho_ka_85},
where G\" ursey model lagrangian was attempted to be written  in a
polynomial form, and recent work,
\cite{mi_93,ba_le_lo_86,le_lo_ba_86,re_00,re_hepth_99}, which
suggested that the gauged form of the NJL model can be interpreted
as a nontrivial theory. In \cite{ho_lu_06} we reinterpreted our
earlier work, \cite{ak_ar_du_ho_ka_pa_82-34}, and showed that
rather than finding a trivial theory, as claimed in
\cite{ar_ho_83}, we ended up in a model where composite particles
took part in physical processes.   The constituent fields,
however, did not interact with each other when perturbation theory
was applied to the model, as already shown in \cite{ar_ho_83}. In
\cite {ho_lu_ta_hepth_06}, we showed that, when this model is
coupled to a constituent $U(1)$ gauge field, we were mimicking a
gauge Higgs-Yukawa (gHY) system, which had the known problems of
the Landau pole, with all of its connotations of triviality.

\noindent The essential point of our earlier work was the fact
that  the propagator of the composite scalar field was equal to
${{\epsilon}\over{p^2}}$.  Since $\epsilon$ goes to zero as the
cutoff is removed, many of the diagrams, where the scalar field
propagator takes part as an internal line, become convergent.  We
could show that there was no breaking of the chiral symmetry, thus
no mass generation, for the fermion fields in our model in higher
orders of perturbation theory.

 \noindent
Here we will study our original model \cite{ho_lu_06}, coupled to
a $SU(N)$ gauge field and use solely  renormalization group
techniques.
We start
with the description of our starting model without the gauge field
\cite {ho_lu_06}.  Then we derive the renormalization group equations
 (RGEs) in one loop, and try to
derive the criteria for obtaining non-trivial fixed points for the
coupling constants of the theory.  Here we closely follow the line
of discussion followed in our  reference \cite{ha_ki_ku_na_94}. In
our model, however, there is a composite scalar field with a
propagator completely different from a constituent scalar field
used in this reference.  This gives rise to RGE's in our case
which are different from those given by Harada et al.  Since our
starting models are different the motivation of our work  is
different from that of this reference. We show that the
renormalization group equations  point to   the non-triviality of
the model when it is coupled to an $SU(N)$ gauge field.  We end up
with a few remarks in the last section. %
\pagebreak
\section{The Original Model}
Our initial model is
given by the Lagrangian

\begin{equation}
L = {i\overline{\psi}} \partial \!\!\!/ \psi + g {\overline{\psi}}
\psi \phi +\xi ( g{\overline {\psi}} \psi -a\phi^{3} ).\label{cl}
\end{equation}

\noindent Here the only terms with kinetic part are the spinors.
$\xi$ is a Lagrange multiplier field, $\phi $ is a scalar field
with no kinetic part, $g$ and $a$ are coupling constants. This
expression contains two constraint equations, obtained from
writing the Euler-Lagrange equations for the $\xi$ and $\phi$
fields.  Hence, it should be quantized by using the Dirac
constraint analysis as performed in reference \cite{ho_lu_06}.

\noindent
The Lagrangian given above is just an attempt in writing the
original G\" ursey Lagrangian

\begin{equation}
 L={i\overline{\psi}} \partial \!\!\!/ \psi + g' ({\overline{\psi}} \psi)^{4/3}
 ,\label{gl}
\end{equation}
in a polynomial form.

\noindent We already showed how  the $\gamma^{5} $ invariance of
the  G\"{u}rsey  Lagrangian  , which  prevents the fermion field
from acquiring a finite mass in higher orders, is retained in  our
model,  and the fact that our model is equivalent to the original
G\"{u}rsey model only classically  in  \cite{ho_lu_06}.

\noindent
To quantize the latter system consistently we proceed via the path
integral method. This procedure is carried out in reference
\cite{ho_lu_06}.  At the end of these calculations we find out
that we can write the constrained lagrangian given in equation
(\ref{cl}) as

\begin{equation}
L' = {i\overline{\psi}}[\partial \!\!\!/  +ig
\Phi]\psi-{{a}\over{4}}(\Phi^{4}+2\Phi^{3}\Xi-2\Phi\Xi^{3}-
\Xi^{4})+{i\over{4}}c^*(\Phi^{2}+2\Phi\Xi+\Xi^{2}) c, \label{eff1}
\end{equation}
where the effective lagrangian  is expressed in terms of scalar
fields $\Phi$, and $\Xi$, ghost fields  $c $, $c^*$ and spinor
fields only.

\noindent
The fermion propagator is the usual Dirac propagator in lowest
order, as can be seen from the lagrangian.  After integrating over
the fermion fields in the path integral, we obtain the effective
action. The second derivative of the effective action with respect
to the $\Phi$ field gives us the induced inverse propagator for
the $\Phi$ field, with the infinite part given as

\begin{equation} \mbox{inf} \left[ {{ig^2}\over{ (2\pi)^4}} \int {{d^4
p}\over {p\!\!\!/(p\!\!\!/+q\!\!\!/)}}\right]=
 {{g^2  q^2}\over {4\pi \epsilon}}.
\end{equation}
Here dimensional regularization is used for the momentum integral
and $\epsilon = 4-n$.  We see that the $\Phi$ field propagates as
a massless field.

\noindent
When we study the propagators for the other fields, we see that no
linear or quadratic term in $\Xi$ exists, so the one loop
contribution to the $\Xi$ propagator is absent. Similarly the
mixed derivatives of the effective action with respect to $\Xi$
and $\Phi$  are zero at one loop, so no mixing between these two
fields occurs. We can also set the propagators of the ghost fields
to zero, since they give no contribution in the one loop
approximation.  The higher loop contributions are absent  for
these fields.

In reference \cite{ho_lu_06} we also studied the contributions to
the fermion propagator at higher orders and we found, by studying
the Dyson-Schwinger equations for the two point function, that
there were no new contributions.  We had at least one phase where
the mass of the spinor field was zero.

\section{Renormalization Group Equations}

\noindent Here we couple an $SU(N_C)$ gauge field to the model. We
also take spinors with different flavors,  up to $N_f$.  The new
lagrangian reads:

\begin{equation}
L = {\sum_{i=1}^{N_{f}}i\overline{\psi}_{i}} D \!\!\!\!/ \psi_{i}
+ g \sum_{i=1}^{N_{f}}{\overline{\psi}_{i}} \psi_{i} \phi +\xi
\left(g\sum_{i=1}^{N_{f}} {\overline {\psi}_{i}} \psi_{i}
-a\phi^{3} \right)-\frac{1}{4} Tr [F_{\mu \nu} F^{\mu \nu}].
\label{newlag}
\end{equation}

Upon performing constraint analysis similar to the one performed
in \cite{ho_lu_06}, we see that we have to satisfy
\begin{eqnarray}
 \sum_{i=1}^{N_f} \overline{\psi}_i \psi_i -a\phi^3 =0, \hspace{2cm}
 3 a \xi \phi^2-g \sum_{i=1}^{N_f} \overline{\psi}_i \psi_i =0.
\end{eqnarray}
After calculating the constraint matrix,  raising the result to
the exponential by using ghost field,  and performing the
transformations $\Phi = \phi + \xi $ and $ \Xi= \phi - \xi$ we get
similar equations as given in equation (\ref{eff1}). We see that
both the $\Xi$ and the ghost fields coming from the compositeness
constraint decouple from our model.

At this point we have to note that there are two kinds of ghost
contributions in the new model. The ghosts coming from the gauge
condition on the vector field do not decouple, and contribute to
the renormalization group equations in the usual way.  We impose
these constraints on equation (\ref{newlag}).

After these steps we start with the effective lagrangian given as

\begin{equation}
L'' = -\frac{1}{4} Tr [F_{\mu \nu} F^{\mu
\nu}]-\frac{a}{4}\Phi^4+\sum_{i=1}^{N_{f}}\overline{\psi}_{i}
iD\!\!\!\!/\psi_{i}-\sum_{i=1}^{N_{f}}g\Phi\overline{\psi}_{i}\psi_{i}
+ L_{\mbox{ghost}}+ L_{\mbox{gauge fixing}}.
\end{equation}
Here $N_f$ is the number of flavors.  The gauge field belongs to
the adjoint representation of the color group $SU(N_C) $ where
$D_{\mu}$ is the color covariant derivative. $g,a,e$ are the
Yukawa, quartic scalar and gauge coupling constants, respectively.
We take $N_{f}$ in the same order as $N_{C}$.

\noindent
In the one loop approximation, the
renormalization group equations read as

\beq 16\pi^2\frac{d}{dt}e(t) &=& -be^3(t),\label{yenimodelErge}\\
     16\pi^2\frac{d}{dt}g(t) &=& -cg(t)e^2(t),\label{yenimodelGrge}\\
     16\pi^2\frac{d}{dt}a(t) &=& -ug^4(t) ,\label{yenimodelArge}
\eeq%

where $b$,$c$ and $u$ are positive constants given as

\begin{equation}
b=\frac{11N_C-4T(R)N_f}{3}, \hspace{1cm} c=6C_2(R), \hspace{1cm}
u=8N_fN_C . \end{equation} Here $C_2(R)$ is a second Casimir,
$C_2(R)=\frac{(N_{C}^{2}-1)}{2N_{C}}$ and $R$ is the fundamental
representation with $T(R)=\frac{1}{2}$. We take $\mu_0$ as a
reference scale at low energies, $t=ln (\mu/\mu_0)$, where $\mu$
is the renormalization point.

\noindent
 In the RGE we see that the diagrams, where scalar
propagators take part, are down by powers of $\epsilon$. Hence we
do not have contributions proportional to $a^2(t)$, $g^3(t)$ and
$a(t)g^2(t)$, as one would have in the gHY system as described in
the work of  \cite{ha_ki_ku_na_94}. Since the diagrams, omitted in
\cite{ha_ki_ku_na_94} via a $\frac {1}{N_c}$ analysis, are down by
an order of $\epsilon $ in our analysis, we do not need  a
relation between $ N_C, N_f$  and the coupling constants at this
point.
\subsection{Solutions of the RGE's}

The solution for the first RG equation (\ref{yenimodelErge}) can
be obtained easily as

\beq e^{2}(t)=e_{0}^{2}\Bigg(1+\frac{b\alpha_0}{2\pi}t\Bigg)^{-1},
\label{e_nin_cozumu}\eeq%
where $\alpha_0=\frac{e_{0}^2}{4\pi}$. Define
\beq
\eta(t)\equiv\frac{\alpha(t)}{\alpha_{0}}\equiv\frac{e^2(t)}{e_0^2},
\eeq
where $e_0=e(t=0)$ which is the initial value at the reference
scale $\mu_0$. For the solution of the second RG equation
(\ref{yenimodelGrge}) we can define a RG invariant $H(t)$ as

\beq H(t)=(c-b)\eta^{-1+\frac{c}{b}}(t)\frac{e^2(t)}{g^2(t)}. \eeq
Since $H(t)$ is a constant, we  call it $H_0$. Then, the solution
of the gauge coupling constant can be written as

\beq g^2(t)=\frac{(c-b)e^2_0}{H_0}\eta^{\frac{c}{b}}(t).
\label{g_nin_cozumu}\eeq The solution of the last RG equation
(\ref{yenimodelArge}) can be defined by another RG invariant
$K(t)$,  given as

\beq K(t)=-u\eta^{-1+\frac{2c}{b}}(t)\Bigg[
1-\frac{2(2c-b)}{u}\frac{a(t)}{g^2(t)}\frac{e^2(t)}{g^2(t)}\Bigg]
.\eeq We can then write

\beq a(t)=\frac{u}{2(2c-b)}\frac{g^2(t)}{e^2(t)}g^2(t)
\bigg[1+\frac{K_0}{u}\eta^{1-\frac{2c}{b}}(t)\bigg]
\label{a(t1)}.\eeq Here $K_0$ is the value of the RG invariant. We
can rewrite equation (\ref{a(t1)}) as

\beq a(t)=\frac{u(c-b)^2e_0^2}{2H_0^2(2c-b)}
\bigg[\eta^{-1+\frac{2c}{b}}(t)+\frac{K_0}{u}\bigg]
\label{a(t2)}.\eeq When we check the ultraviolet limit now, we
find

\beq \eta(t \rightarrow\infty)\rightarrow
\ %
\begin{array}{ll}
  +0, \hspace{5mm}& \hbox{$b>0$;} \\
  \end{array}%
\eeq

\beq \eta^{\frac{c}{b}}(t \rightarrow\infty)\rightarrow \ %
 \begin{array}{ll}
  +0, \hspace{5mm}& \hbox{$c,b>0$;} \\
  \end{array}%
\eeq and \beq \eta^{-1+\frac{2c}{b}}(t
\rightarrow\infty)\rightarrow
\ \left\{%
\begin{array}{lll}
 +0, \hspace{5mm}& \hbox{$2c>b$;} \\
 +0, \hspace{5mm}& \hbox{$2c>b>c$;} \\
 +\infty, \hspace{5mm}&  \hbox{$b>2c$.} \\
\end{array}%
\right.\eeq

\noindent We see that the constants $H_0$ and $K_0$ play important
roles on the behavior of solutions of coupling equations
(\ref{e_nin_cozumu}),(\ref{g_nin_cozumu}),(\ref{a(t2)}). For
$c>b$, $H_0$ should be positive; for $c<b$, $H_0$ should be
negative to have  the Yukawa coupling take a real value.  This is
necessary to have a unitary theory.  Also for a region $c<b<2c$,
with $H_0<0$, the unitarity condition is satisfied for all
coupling constants. $K_0\geq0$ condition is also needed for
stability of the vacuum.  If $K_0 <0$, we get
$a(t\rightarrow\infty)<0$ , which raises the problem of the vacuum
instability.

\noindent
Next we study the different limits our parameters can take:

\subsubsection{$b\rightarrow+0$ limit case for finite $t$}

We find

\beq e^2(t)&=&e^2_0, \nonu \\
g^2(t)&=&\frac{ce^2_0}{H_0}exp(-\frac{\alpha}{\alpha_c}t), \\
a(t)&=&\frac{uce^2_0}{4H^2_0}\bigg[exp(-\frac{2\alpha}
{\alpha_c}t)+\frac{K_0}{u}\bigg].\nonu \eeq Here
$\frac{c}{2\pi}=\frac{1}{\alpha_c}$ and $\alpha_0=\alpha$. This
means that when we set the $b$ term  to zero, the Yukawa running
coupling constant decreases exponentially to zero.  For this limit
the gauge and the quadratic coupling constants go just to a
constant.

\subsubsection{$c\rightarrow b$ limit case for finite $t$}

If $c$ approaches $b$, the limit depends on the value of $H_0$. If
$H_0$ is non zero, $g^2(t)$ goes to zero.
If $H_0$ goes to zero as a constant times $c-b$, i.e.
$H_0=\frac{c-b}{H_1}$, we find that $g^2(t)$ and $a(t)$ are both
proportional to $e^2(t)$ as follows

\beq g^2(t)=H_1 e^2(t), \hspace{3mm} H_1>0;\eeq

\beq a(t)=\frac{ue_0^2H_1^2}{2b}\bigg[\eta(t)+\frac{K_0}{u}\bigg].
\label{h1}\eeq

\subsubsection{$2c\rightarrow b$ limit case for finite $t$}

When $2c$ approaches $b$, the behavior of $a(t)$ changes. If we
set $ \frac{K_0}{u}=-1+\frac{2c-b}{b}K_1$, then $a(t)$ goes as $ln
\eta(t)$

\beq a(t)=\frac{ube_0^2}{8H_0^2}\bigg[K_1+ln \eta(t)\bigg].\eeq
This behaviour is not allowed since $a(t)$ diverges as
$t\rightarrow+\infty$.

\section{Non-triviality of the system}

In this section we  use the preceding results to investigate
the non-triviality of the system with several criteria such as:

\noindent
 All the running coupling constants:

\begin{itemize}
    \item should not diverge at finite t$>$0 (no Landau poles);
    \item should not vanish identically;
    \item should not violate the consistency of the theory such as
    unitarity and/or vacuum stability.
\end{itemize}%

\noindent
Since the composite scalar field is the novel feature of our
model, we will not consider the case when the scalar field is
completely decoupled from the theory.

\subsection{Fixed Point Solution}

We derive the expressions given below from the RGE equations.

\beq 8\pi^2\frac{d}{dt}\Bigg[\frac{g^2(t)}{e^2(t)}\Bigg]=(b-c)
\Bigg[\frac{g^2(t)}{e^2(t)}\Bigg]e^2(t), \label{c1}\eeq

\beq
8\pi^2\frac{d}{dt}\Bigg[\frac{e^2(t)}{g^2(t)}\frac{a(t)}{g^2(t)}\Bigg]=(2c-b)
\Bigg[\frac{e^2(t)}{g^2(t)}\frac{a(t)}{g^2(t)}-\frac{u}{2(2c-b)}\Bigg]e^2(t)
\label{c2}.\eeq%
For the fixed point solution, $b$ equals $c$ in equation
(\ref{c1}). For this value, there is a single solution which
satisfies both equations (\ref{c1}) and (\ref{c2}). This solution
is given as,

\beq \frac{e^2(t)}{g^2(t)}=\frac{1}{H_{1}}, \eeq
where $H_1$ is a constant, and %

\beq\frac{a(t)}{g^2(t)}=\frac{uH_1}{2c}.\eeq
If we take $H_0=H_1(c-b)$ approaching zero as $c$ approaches to
$b$, while $K_0=0$ in equation (\ref{a(t2)}) , then we find

\beq g^2(t)=H_1e^2(t) ,\eeq \beq a(t)=\frac{uH_1}{2c}g^2(t) .\eeq
Since $\frac{g^2(t)}{e^2(t)}$ and $\frac{a(t)}{g^2(t)}$ are
constants, the behavior of the Yukawa and quartic scalar couplings
are completely determined by the gauge coupling. This corresponds
to "coupling constant reduction" in the sense of Kubo, Sibold and
Zimmermann \cite{ku_si_zi_89}. In the context of the RGE, it
corresponds to the Pendleton-Ross fixed point \cite{pe_ro_81}.

%

\subsection{Yukawa Coupling}
As seen from the previous sections the behavior of the Yukawa
coupling depends on whether $c>b$ or $c<b$. The point where  $c=b$
needs a special care. Moreover the sign of the $H_0$ is important.

\subsubsection{c$>$b case}

In this case $H_0$ should not  equal to zero. Then we find in the
UV limits

\beq g^2(t\rightarrow\infty)\rightarrow\ \left\{%
\begin{array}{ll}
    +0, \hspace{5mm}& \hbox{$H_0>0$;} \\
    -0, \hspace{5mm}& \hbox{$H_0<0$.} \\
\end{array}%
\right.     \eeq So the Yukawa coupling is asymptotically free. As
it is seen, the sign of the RG invariant is important. It should
be chosen positive not to cause the violation of stability of the
vacuum.

\noindent
In Figure 1 we plot $g^2$ vs. $e^2$ for $c=8$, $b=7$.  Both
coupling constants approach the origin as $t$ goes to infinity.
Thus, our model fulfills the condition required by the asymptotic
freedom criterion.

\begin{figure}[htb!]
\epsfxsize=95mm \epsffile{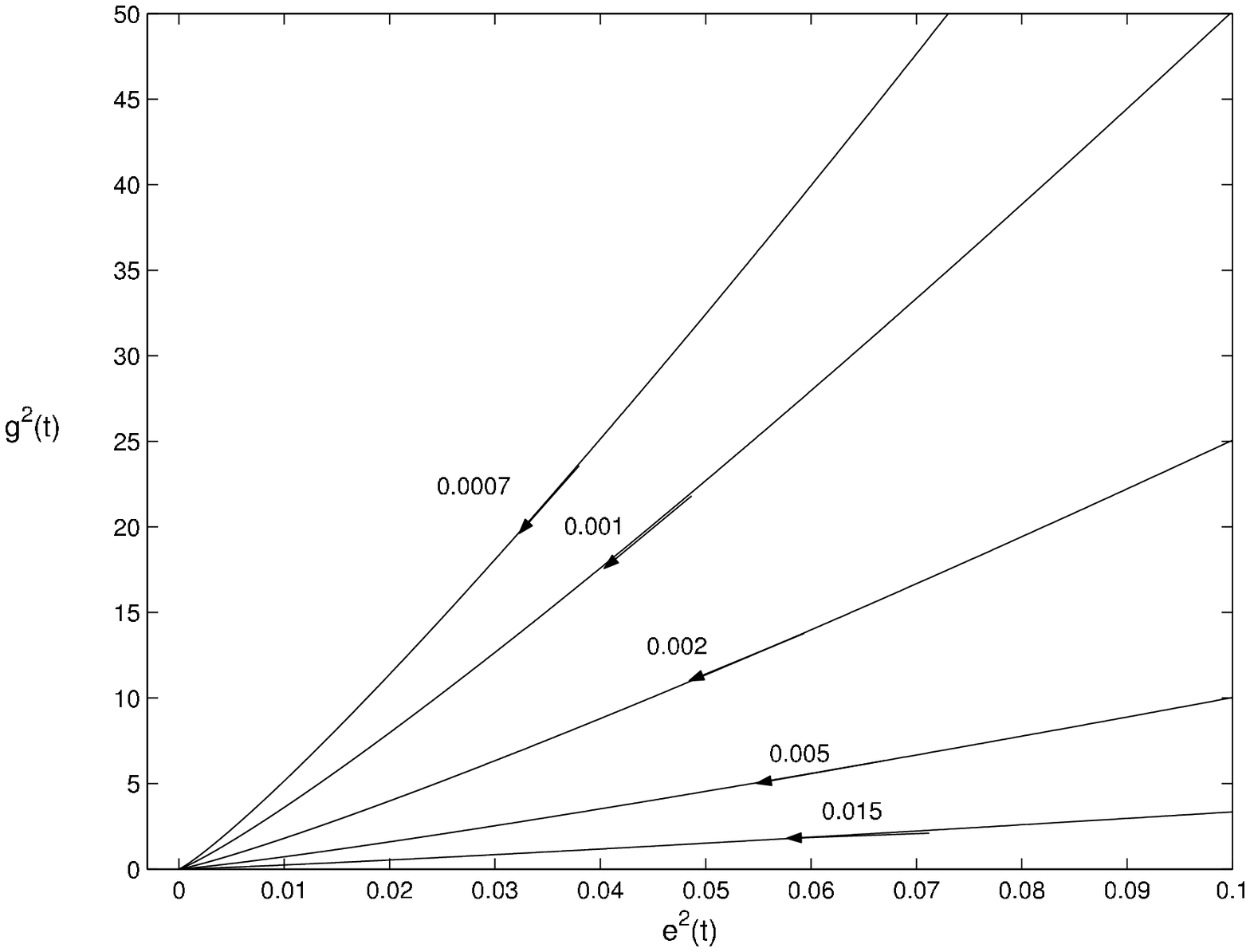} \\
\caption{Plot of $g^2(t)$ vs. $e^2(t)$ for different values of
$H_0$. The arrows denote the flow directions toward the UV
region.} \label{e_kare_g_kare} \end{figure}

\subsubsection{c$<$b case}
In this case with a non zero value of $H_0$
\beq g^2(t\rightarrow\infty)\rightarrow\ \left\{%
\begin{array}{ll}
    -0, \hspace{5mm}& \hbox{$H_0>0$;} \\
    +0, \hspace{5mm}& \hbox{$H_0<0$.} \\
\end{array}%
\right.     \eeq For $H_0<0$, our system satisfies the asymptotic
freedom condition. Our system does not have a Landau pole. In this
respect it differs from the gHY system \cite{ha_ki_ku_na_94}. As
shown below, there is a restriction on the value of $b$ in this
case. \subsubsection{c$=$b case}

This is the fixed point solution analyzed above.

 \beq
g^2(t)=H_1e^2(t).\eeq%


\subsection{Quartic Scalar Coupling}
$a(t)$ can be analyzed with four non trivial limits of the Yukawa
coupling. \begin{itemize}
    \item $c>b$ with $H_0>0$,
    \item $c<b<2c$ with $H_0<0$,
    \item $b>2c$ with $H_0<0$,
    \item $c=b$ with $H_0=0$.
\end{itemize}

\noindent
For $c>b$ case, we should have $H_0>0$,  whereas in the  $c<b<2c$
case we have $H_0<0$. In both cases $K_0$ should be greater or
equal to zero for the stability of the vacuum. In the third case,
$b>2c$ with $H_0<0$, for all the real values of $K_0$, $a(t)$
diverges in the UV limit. This means that there is no chance for a
nontrivial theory in that region. Finally the $c=b$ case with
$H_0=0$ have already be shown in equation (\ref{h1}). It is clear
that in the UV limits $K_0$ should not take negative values.

\noindent
As seen above these constraints give different relations between
numbers of color and flavor. Note that in all the cases studied,
if we take $K_0<0$, one can deduce from  equation (\ref{a(t2)})
that  $a(t)$ can be made equal to zero for a finite value of $t$,
a  situation which should not be allowed. Therefore, we can use only
the option  with $K_0\geq0$.  The standard model with three colors
and six flavors satisfies the $c>b$ case.

\noindent
For $K_0=0$ at the UV limit, the equation (\ref{a(t2)})

\beq a(t)=\frac{u(c-b)^2e_0^2}{2H_0^2(2c-b)}
\eta^{-1+\frac{2c}{b}}(t) \rightarrow +0, \eeq
shows that  the coupling constant is asymptotically free. Also for
a non zero $K_0$, we find in the UV limit

\beq a(t)\rightarrow \frac{(c-b)^2e_0^2K_0}{2H_0^2(2c-b)}. \eeq
Then the sign of the $K_0$ is crucial for the stability of the
vacuum.

\noindent Although for $K_0 > 0$ we do not violate unitarity, we
see that the asymptotic freedom criterion is not satisfied.  The
requirement of this criterion fixes $K_0$ at the value zero. In
Figure 2, we plot the RG flows in $(a(t),g^2(t))$ plane for
different values of $H_0$ higher than zero while the gauge
coupling $\alpha(t=0)$ is fixed to $1$. The origin is the limit
where $t$ goes to infinity, there both coupling constants approach
zero
when $K_0=0$. %

\begin{figure}[htb!]
\epsfxsize=95mm%
\epsffile{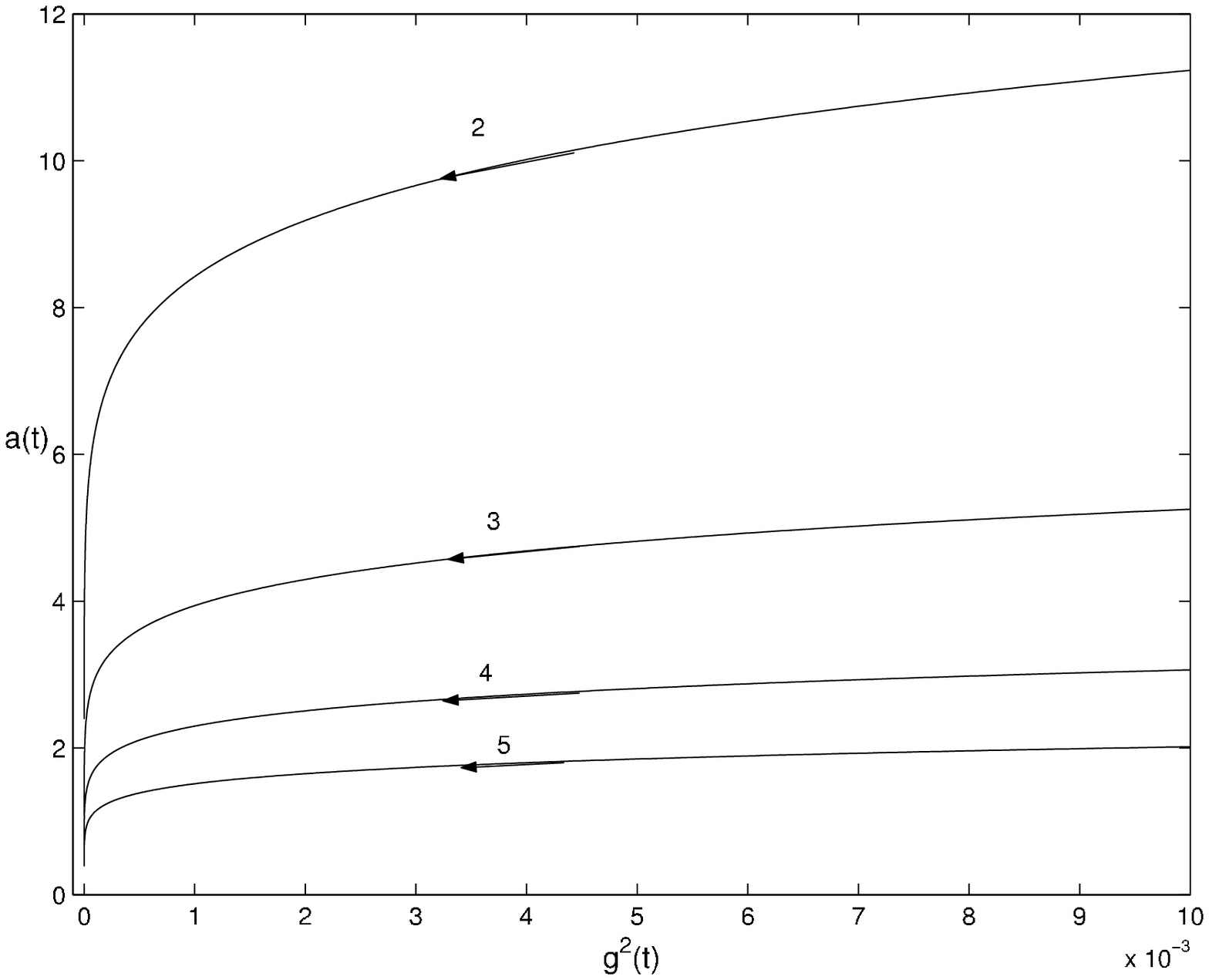}%
\caption{Plot of $a(t)$ vs. $g^2(t)$ for different values of $H_0$
while $K_0=0.$} \label{g_kare_a} \end{figure}

\section{Conclusion}

Here we write the $SU(N)$ gauge version of the polynomial
lagrangian inspired by the G\" ursey model.  In \cite{ho_lu_06} we
had found an interacting model, where only the composites take
part in scattering processes, if only perturbative calculations
are done. Gauging it with a constituent $U(1)$ field resulted in a
model which looked like the gHY system,  with all the problems
associated with the Landau pole \cite{ho_lu_ta_hepth_06}.  When a
$SU(N)$ gauge field is coupled, instead, we find that the
renormalization group equations for the three coupling constants
indicate that this model is nontrivial. All the coupling constants
go to zero asymptotically as the cutoff parameter goes to
infinity, exhibiting the behavior dictated by asymptotic freedom.

\noindent
In equations (\ref{c1}) and (\ref{c2}) we give the equations for
the ratios of the coupling constants and find the fixed points. We
see that we can have nontrivial fixed points.

\noindent
One can apply the exact renormalization group to our model and
obtain the additional vertices as given in our references
\cite{ha_ki_ku_na_94} and \cite{ao_mo_su_te_to_99}. This will be
pursued in the future.

\vspace{5mm}\textbf{Acknowledgement}: We thank Ferhat Ta\c sk\i n
for many discussions while preparing this manuscript.  This work
is supported by the  ITU BAP project no: 31595. The work of M.H.
is supported by TUBA, the Academy of Sciences of Turkey. This work
is also supported by TUBITAK, the Scientific and Technological
Council of Turkey.

\end{document}